\documentclass[prd,superscriptaddress,floatfix,nofootinbib,twocolumn]{revtex4-2}
\pdfoutput=1 
\usepackage[T1]{fontenc} 
\usepackage{amsmath,amssymb,amsfonts,amsthm}
\usepackage{graphicx}
\usepackage[usenames,dvipsnames]{color}
\usepackage{enumitem}
\usepackage{verbatim}
\usepackage[percent]{overpic}
\usepackage{rotating}
\usepackage[colorlinks=true]{hyperref}
\usepackage{array}
\usepackage{mathtools}
\usepackage{lmodern}
\usepackage[normalem]{ulem}
\usepackage{braket}
\usepackage{tensor}
\usepackage{dsfont}
\usepackage{bm}
\usepackage{tikz}

\usepackage{mathrsfs}

\usetikzlibrary{decorations.pathmorphing}
\usepackage{CJKutf8}

\definecolor{patriarch}{rgb}{0.5, 0.0, 0.5}

\newcommand{\x}{\mathsf{x}}

\usetikzlibrary{positioning}
\usetikzlibrary{calc,through,backgrounds}

\theoremstyle{definition}
\newtheorem{definition}{Definition}[section]
\newtheorem{theorem}{Theorem}[section]

\newcommand{\sgn}{\text{sgn}}
\newcommand{\kako}[1]{\left( #1 \right)}
\newcommand{\kagikako}[1]{\left[ #1 \right]}
\newcommand{\ts}[1]{ _{\text{#1}} }

\allowdisplaybreaks

\DeclareMathOperator{\Tr}{Tr}

\newcommand{\dd}{\text{d}}

\newcommand{\sx}{\mathsf{x}}

\newcommand{\ii}{\mathsf{i}}

\hypersetup{
	colorlinks=true,
	linkcolor= black,
	filecolor=magenta,      
	urlcolor=cyan,
	pdftitle={Overleaf Example},
	pdfpagemode=FullScreen,
	citecolor=black
}

\usepackage{float}

\begin{document}
\title{Quenched Entanglement Harvesting}

\author{Adrian Lopez-Raven}
\email{adrian.lopez@uwaterloo.ca}
\affiliation{Department of Physics and Astronomy, University of Waterloo, Waterloo, Ontario, N2L 3G1, Canada}
\affiliation{Perimeter Institute for Theoretical Physics,  Waterloo, Ontario, N2L 2Y5, Canada}

\author{Robert B. Mann}
\email{rbmann@uwaterloo.ca}
\affiliation{Department of Physics and Astronomy, University of Waterloo, Waterloo, Ontario, N2L 3G1, Canada}
\affiliation{Perimeter Institute for Theoretical Physics,  Waterloo, Ontario, N2L 2Y5, Canada}
\affiliation{Institute for Quantum Computing, University of Waterloo, Waterloo, Ontario, N2L 3G1, Canada}

\author{Jorma Louko}
\email{jorma.louko@nottingham.ac.uk}
\affiliation{School of Mathematical Sciences, University of Nottingham, Nottingham NG7 2RD, UK}

\date{June 2025; updated September 2025.\\ aaPublished in Phys.\ Rev.\ D \textbf{112}, 085001 (2025), doi.org/10.1103/lyhy-ftxz.\\ aaFor Open Access purposes, this Author Accepted Manuscript is made available under CC BY public copyright.}

\begin{abstract}
Ultracold fermionic atoms in an optical lattice, with a sudden position-dependent change (a quench) in the effective dispersion relation, have been proposed by Rodr\'iguez-Laguna \textit{et al.\ }as an analogue spacetime test of the Unruh effect. We provide new support for this analogue by analysing the entanglement of a scalar field in a $(1 + 1)$-dimensional continuum spacetime with a similar quench, and the harvesting of this entanglement by a pair of Unruh-DeWitt detectors. We present numerical evidence that the concurrence and mutual information harvested by the detectors are qualitatively similar to those in Rindler spacetime, but they exhibit a small yet noticeable variation when the energy pulse created by the quench crosses the detectors. These findings provide further motivation to implement the experimental proposal of Rodr\'iguez-Laguna \textit{et al.}
\end{abstract}

\maketitle

\section{Introduction}

It has long been known that the vacuum state of a quantum field contains entanglement \cite{summers1985bell, summers1987bell}. This has led to the expectation that such entanglement can be extracted, or harvested, by physical devices
or objects such as atoms \cite{Valentini1991nonlocalcorr, reznik2003entanglement, reznik2005violating}, and that this can take place  even when these objects are spacelike separated.

The general paradigm for the entanglement harvesting protocol \cite{pozas2015harvesting} considers 
two initially uncorrelated detectors that locally interact with a quantum field
in some state (typically the vacuum state).
The amount of harvested correlations is sensitive to both the composition and states of the detectors (for example, their motion) \cite{Doukas.orbit.PhysRevA.81.062320, salton2015acceleration, Zhang.harvesting.circular, Liu.harvesting.reflecting.boundary, Liu:2021dnl, FooSuperpositionTrajectory, Diki.inertial, Liu.acceleration.vs.thermal, ManarKenMutual} as well as 
the  spacetime background \cite{Steeg2009, Cliche.harvesting.weakgrav, smith2016topology, kukita2017harvesting, henderson2018harvesting, ng2018AdS, henderson2019entangling, cong2020horizon, robbins2020entanglement, Xu.Grav.waves.PRD.102.065019, Tjoa2020vaidya, Ken.Freefall.PhysRevD.104.025001, FinnShockwave, Kendra.BTZ, Henderson:2022oyd, caribe2023lensing}. 

This effect is extraordinarily difficult to measure in vacuum, particularly  given the length and time scales involved.  This has motivated efforts to consider alternative settings in which entanglement harvesting can be observed, with
recent experiments detecting correlations of the electromagnetic ground state in a ZnTe crystal  \cite{Benea.Electric.correlation.experiment, Settembrini.Detection.correlation.experiment.2022, Lindel2023separately.experiment} providing further impetus to this end. Indeed there has been a recent proposal to extract entanglement from the ZnTe crystal  \cite{Lindel:2023rfi}, as well as from  quantum surface fluctuations of a Bose-Einstein condensate~\cite{Gooding:2023xxl}.

New analogue settings that behave as relativistic quantum field theories, originally proposed in the context of testing  the Unruh Effect, provide a promising avenue in which to test entanglement harvesting. These have a causal structure that is bounded by the speed of sound of the medium and not the speed of light, allowing for entanglement harvesting properties to be within reach of measurement techniques. 

Amongst the various analog settings, one of particular note is that proposed by   Rodríguez-Laguna \textit{et al}  \cite{Rodriguez_Laguna_2017} and Kosior \textit{et al} \cite{Kosior_2018}, in which an optical lattice with ultracold fermionic atoms is used to model a Dirac field in a curved spacetime in a  setup that permits control of the low energy effective Hamiltonian of the system. To model the Unruh effect, the Hamiltonian is first set  to be that  of a free Dirac fermion in Minkowski spacetime. Then there is a ``quench'' of the Hamiltonian, so that it 
effectively 
becomes the one associated with a free fermion in a spacetime whose $(1+1)$-dimensional part is
\begin{align}
ds^{2} = -\chi^{2}d\eta^2 + d\chi^2 , 
\label{eq:rindlermetric}
\end{align}
known as the Rindler metric.

In this proposal, the field is initialized in its Minkowski vacuum state, after which it  evolves through the quench~\cite{Rodriguez_Laguna_2017}. Numerical simulations indicated  that Unruh-like effects should be present, despite the analog nature of the proposed experiment.  It was subsequently  argued in a different way that Unruh like effects should be present in these lattice setups~\cite{Louko_2018}. 
Specifically, it was shown that a free massless scalar field in the $(1+1)$-dimensional  
quench spacetime (described in detail below in Section~\ref{sec:the_quench_spacetime}) experiences an Unruh effect in the post-quench region. This provides further evidence 
that Unruh-like effects should be observable in ultracold-fermionic-atom lattices. 

Here we study the \emph{entanglement} properties of a massless scalar in this $(1+1)$-dimensional quench spacetime. In particular, we investigate the entanglement  harvesting properties of a pair of accelerated 
Unruh-DeWitt (UDW) detectors \cite{Unruh1979evaporation,DeWitt1979} 
in this setup. 
We find that indeed the harvesting phenomenon is present, and exhibits similar properties to the entanglement harvested by UDW detectors on accelerated trajectories in flat Minkowski spacetime.  Additionally, we show that the entanglement harvested exhibits a small but distinct dependence on the energy pulse that the quench creates in the post-quench part of the spacetime.

Our results strengthen the expectation that entanglement harvesting should also be present in analog setups   \cite{Rodriguez_Laguna_2017}, \cite{Kosior_2018}, providing further motivation to realize their experimental proposals.

The structure of our paper is as follows.  In Section \ref{sec:the_quench_spacetime} we provide a brief description of the quench spacetime and the stress energy tensor of a quantised massless scalar field therein.
In Section \ref{sec:Ingredients_of_Ent_Harv} we describe the Wightman function, trajectories, and switching functions used to calculate the entanglement harvested. In Section \ref{sec:setup_of_calculation}, we describe in detail the specific parameter ranges over which the entanglement was calculated. In Section \ref{sec:results} we show the obtained entanglement harvesting results, making some brief remarks about them, and finally in Section \ref{sec:conclusions} we report our conclusions of the work. In Appendix~\ref{sec:about_code}, we briefly discuss the code used for the numerical entanglement calculations and outline the validation checks taken to ensure the robustness of the results.

\section{The Quench spacetime}\label{sec:the_quench_spacetime}

The quench spacetime  \cite{Louko_2018} consists of a flat spacetime in the past of a distinguished spacelike hypersurface, and of a regularized double Rindler spacetime in the future of this hypersurface. The metric reads 
\begin{align}\label{eq:quench_metric}
ds^2 = 
\begin{cases}
	-(\chi^2 + b^2)d\eta ^2 + d\chi^2 & \text{for } \eta \geq 0\\
	-dt^2 + dx^2 & \text{for } t < 0
\end{cases}
\end{align}
using Rindler-like coordinates in the future half, where $b$ is a positive parameter, and Minkowski coordinates in the past half. 
The quench occurs at $t=0=\eta$, where the halves are joined so that $x=\chi$. 
In the future half, the
metric may be expressed in the conformally flat form
\begin{equation}
	ds^2 = b^2 \cosh^2(y) (-d\eta^2 + dy^2)
\end{equation}
through the coordinate transformation
\begin{equation}\label{eq:y_coord_definition}
	\chi = b \sinh(y) . 
\end{equation}

This spacetime has several features that make it a good candidate for modelling the setup proposed in \cite{Rodriguez_Laguna_2017}. Apart from the obvious point that it models rapid transition between flat Minkowski and a Rindler like spacetime (which is 
the desired situation for experiment \cite{Rodriguez_Laguna_2017,Kosior_2018}),  it possesses two additional properties that make it resemble the proposed setup:
\begin{itemize}
\item  
In the optical lattice case, it is expected that the $ \chi = 0$ singularity in \eqref{eq:rindlermetric}
gets removed by the discrete nature of the lattice. In the quench spacetime~\eqref{eq:quench_metric}, the singularity has been regularized by the deformation introduced by the $b$ parameter.
\item In the optical lattice, it is expected that there are no Rindler wedges bordered by Killing horizons and separated by future/past causal diamonds. In the quench spacetime it is also the case that those features are not present, since the spacetime is geodesically complete, under a reasonable interpretation of a `geodesic' across the metric discontinuity at $\eta=0=t$.
\end{itemize}

For a quantum massless scalar field on the quench spacetime, 
prepared in the Minkowski vacuum in the pre-quench region, the Wightman function in the post-quench region may be expressed in terms of elementary functions~\cite{Louko_2018}. This makes it possible to compute in the post-quench region quantum observables in terms of elementary functions. 

In the post-quench region, $\eta>0$, the stress-energy tensor (SET) is given by \cite{Louko_2018} 
\begin{align}
	& T_{\eta \eta}=\frac{1}{8 \pi \cosh^2 \! y}-\frac{1}{16 \pi}\left(\frac{1}{\cosh^2(\eta-y)}+\frac{1}{\cosh ^2(\eta+y)}\right) \\
	& T_{y y}=\frac{1}{24 \pi \cosh^2 \! y}-\frac{1}{16 \pi}\left(\frac{1}{\cosh ^2(\eta-y)}+\frac{1}{\cosh ^2(\eta+y)}\right) \\
	& T_{\eta y}=\frac{1}{16 \pi}\left(\frac{1}{\cosh ^2(\eta-y)}-\frac{1}{\cosh ^2(\eta+y)}\right)
\end{align}
At the quench, $\eta\to0_+$, the only nonvanishing component of the SET is 
$T_{yy} \to - 1/(12 \pi \cosh^2 \! y)$, showing that the quench initially creates a negative pressure but no energy density. 
In the evolution after the quench, 
$T_{\eta\eta}$ and $T_{yy}$ 
each consist of a positive static contribution, peaked around $y=0$, 
and negative pulses travelling to the left 
and right at the speed of light, 
peaked around the null rays at $y=\pm\eta$. At given~$\eta$, the pulses have a peak of width of order unity in~$y$, and at given $|y|\gtrsim1$, they have a peak of duration of order unity in~$\eta$; the falloff outside the peak is exponential, both in $\eta$ and in~$y$. 
In the late time limit at fixed~$y$, 
the pulses have passed, with 
$T_{\eta\eta} \to 1/(8 \pi \cosh^2 \! y)$, 
$T_{yy} \to 1/(24 \pi \cosh^2 \! y)$ and $T_{\eta y} \to 0$. 
The initial negative pressure hence evolves at late times 
into a static positive energy density and a static positive pressure. 

In what follows, we will consider how these pulses in the SET are correlated with the entanglement harvested by a pair of model particle detectors.

\section{Ingredients for Entanglement Harvesting}\label{sec:Ingredients_of_Ent_Harv}

\subsection{Preliminaries}

We shall model a UDW detector \cite{Unruh1979evaporation,DeWitt1979} as a pointlike qubit having two energy levels, denoted by 
$|g\rangle$ and~$|e\rangle$, 
with the respective energy eigenvalues $0$ and~$\Omega$. 
For $\Omega>0$, $|g\rangle$ is the ground state and $|e\rangle$ is the excited state; 
for $\Omega<0$, the roles of $|g\rangle$ and $|e\rangle$ are reversed.
Since much of the formalism on UDW detectors is well known (see for example \cite{ManarKenMutual,Ken.Freefall.PhysRevD.104.025001} and references therein), 
we shall only review the necessary elements needed for our study.

For the entanglement harvesting protocol to be implemented, at least two detectors are required. Each  is assumed to couple to a massless scalar field $\phi$ 
via the interaction Hamiltonian 
\cite{Aubry2014derivative,Aubry2018Vaidya,Tjoa2020vaidya}
\begin{equation}
H_{\text{int} D}= \lambda_D \zeta_D(\tau_D)\mu_D(\tau_D)\partial_{\tau_D}\phi\bigl(\x_D(\tau_D)\bigr) 
\ , 
\label{eq:HintD}
\end{equation}
where $\x_D(\tau)$ is the trajectory of detector $D \in \{A,B\}$, 
parametrized by its proper time~$\tau_D$, and
$\mu_D(\tau)
=
        \ket{e_D} \bra{g_D} e^{ \ii \Omega_D \tau }
        +
        \ket{g_D} \bra{e_D} e^{ -\ii \Omega_D \tau }$ is 
its monopole moment operator. 
We may refer to detector $A$ as Alice and detector $B$ as Bob. The quantity
$\lambda_D$ is the coupling constant for detector $D$ and $\zeta_D$ is a switching function, specifying how detector $D$ is switched on and off.  
The total interaction Hamiltonian is
\begin{align}
    \hat H\ts{I}^t(t)
    &= 
    \dfrac{\dd \tau\ts{A}}{\dd t}
    \hat H\ts{A}^{\tau\ts{A}}(\tau\ts{A}(t)) 
    +
    \dfrac{\dd \tau\ts{B}}{\dd t}
    \hat H\ts{B}^{\tau\ts{B}}(\tau\ts{B}(t)) , 
\end{align}
in terms of a common coordinate time~$t$, using  the time-reparametrization property \cite{Pablo2018rqo, Tales2020GRQO}. 
For simplicity, we shall from now take the two detectors to be identical, with
$\Omega_A=\Omega_B=\Omega$, $\lambda_A= \lambda_B$.  

Note that we have employed in \eqref{eq:HintD} the \textit{derivative coupling\/} detector model \cite{Aubry2014derivative,Aubry2018Vaidya,Tjoa2020vaidya}, in which the
detector couples to the field momentum;  the 
interaction Hamiltonian is then linear in the proper time derivative of the field, rather than in the field itself. 
The consequence of this is that the post-interaction reduced density matrix of the two-detector system will not involve the field's Wightman function, but instead the two-point function of the field's derivatives, given by 
\begin{align}\label{eq: derivative-twopoint-distributions}
    \mathcal{A} \big(\sx_i(\tau_i) ,\sx_j(\tau'_j) \big) = \bra{0 }{\partial_{\tau_i} \hat \phi(\sx_i(\tau_i)) \partial_{\tau'_j} \hat \phi(\sx_j(\tau'_j))}\ket{0} , 
\end{align}
where $\ket{0}$ is the state in which the field $\phi$ had been prepared before the interaction. 
The advantage of $\mathcal{A}$ \eqref{eq: derivative-twopoint-distributions} is that it is insensitive to the infrared ambiguity in the Wightman function of a massless scalar field in two spacetime dimensions. Pedagogical discussions of the infrared ambiguity are given in~\cite{Fulling:1987otn,Kay:2000fi}, and a technical discussion in terms of Hadamard renormalisation is given in~\cite{Decanini:2005eg}. 

The time evolution operator of the coupled system is the unitary operator 
\begin{align}
    \hat U\ts{I}&=
    \mathcal{T} 
    \exp
    \kagikako{
        -\ii \int_{-\infty}^\infty \!\!\dd t\,\hat H^t\ts{I} (t)
    } \label{eq:time-evolution operator}\,,
\end{align}
where $\mathcal{T}$ is the time-ordering symbol. 
Taking the density operator of the total system to initially be in the state
\begin{align}
    \rho_0=
    \ket{ g\ts{A} } \bra{ g\ts{A} }
    \otimes \ket{ g\ts{B} } \bra{ g\ts{B} }
    \otimes \ket{0  } \bra{0}\,, 
\end{align}
where $\ket{0}$ is the initial state of the field, 
we obtain the time-evolved density matrix for small $\lambda$
\begin{align}
    \rho
    &= \rho_0 + \sum_{i+j=1}^2 \rho^{(i,j)}
    + O(\lambda^3)\,,
\end{align}
where 
$\rho^{(i,j)}\coloneqq \hat U^{(i)} \rho_0 \hat U^{(j)\dagger} $
and 
\begin{align}
    \hat U\ts{I}=
    \mathds{1} + \hat U^{(1)} + \hat U^{(2)} + O(\lambda^3)\,,
\end{align}
where $\hat U^{(k)}$ is of order $\lambda^k$ given by
\begin{align}
    \hat U^{(1)}&=
    -\ii \int_{-\infty}^\infty \!\!\dd t\,\hat H^t\ts{I}(t), \\
    \hat U^{(2)}&=
    - \int_{-\infty}^\infty \!\!\dd t \int_{-\infty}^t \!\!\dd t'\,\hat H^t\ts{I}(t) \hat H^{t'}\ts{I}(t')\,.
\end{align}
This gives
\begin{align}\label{eq:reduced_density_matrix}
\rho\ts{AB} = 
\begin{pmatrix}
	1 - \mathcal{L}_{AA} - \mathcal{L}_{BB} & 0      & 0      & \mathcal{M}^* \\
	0                   & \mathcal{L}_{BB} & \mathcal{L}_{BA} & 0   \\
	0                   & \mathcal{L}_{AB} & \mathcal{L}_{AA} & 0   \\
	\mathcal{M}                   & 0      & 0      & 0
\end{pmatrix}
\end{align}
for the reduced density matrix of the pair of detectors in the basis $\ket{g\ts{A}} \ket{g\ts{B}}=[1,0,0,0]^\top$, $\ket{g\ts{A}} \ket{e\ts{B}}=[0,1,0,0]^\top$, $\ket{e\ts{A}} \ket{g\ts{B}}=[0,0,1,0]^\top$, and $\ket{e\ts{A}} \ket{e\ts{B}}=[0,0,0,1]^\top$.  The quantities in \eqref{eq:reduced_density_matrix} are
\begin{align}
    &\mathcal{L}_{ij}
    =\lambda^2 
    \int_{-\infty}^\infty \dd \tau_i 
    \int_{-\infty}^\infty \dd \tau_j'\,
    \zeta_i(\tau_i) 
    \zeta_j(\tau_j')
    e^{ -\ii\Omega (\tau_i - \tau_j' ) }\notag\\
    &\hspace{3cm}\times\mathcal{A}(\sx_i(\tau_i) ,\sx_j(\tau_j'))\,, \label{eq:Lij}\\
    &\mathcal{M}=
    -\lambda^2 
    \int_{-\infty}^\infty \dd\tau\ts{A}
    \int_{-\infty}^\infty \dd\tau\ts{B}\,
    \zeta\ts A(\tau\ts{A}) 
    \zeta\ts B(\tau\ts{B})
    e^{ \ii\Omega (\tau\ts{A} + \tau\ts{B}) } \notag \\
    &\hspace{10mm}\times 
    \bigg[
        \Theta \big( t(\tau\ts{A}) - t(\tau\ts{B}) \big)
        \mathcal{A} \big(\sx\ts{A}(\tau\ts{A}), \sx\ts{B}(\tau\ts{B}) \big)\notag\\
        &\hspace{1.2cm}+\Theta \big( t(\tau\ts{B}) - t(\tau\ts{A}) \big)
        \mathcal{A} \big( \sx\ts{B}(\tau\ts{B}) , \sx\ts{A}(\tau\ts{A}) \big)
    \bigg]\,,
    \label{eq:nonlocal M}
\end{align}
where $i,j \in \{A,B\}$, $\Theta(z)$ is the Heaviside step function, and $\mathcal{A}$ is the correlation function~\eqref{eq: derivative-twopoint-distributions}. 
The quantities $\mathcal{L}\ts{AA}$ and $\mathcal{L}\ts{BB}$ are the transition probabilities of Alice and Bob, respectively, whereas $\mathcal{M}$ and $\mathcal{L}\ts{AB}(=\mathcal{L}\ts{BA}^*)$ correspond to   nonlocal terms that depend on both trajectories simultaneously. The quantity $\mathcal{M}$ is responsible for entangling the detectors, whereas $\mathcal{L}\ts{AB}$ is used for calculating the mutual information. 

From these expressions and the discussion above, we see that to calculate the time evolved reduced density matrix~$\rho\ts{AB}$, we need only to determine the $\mathcal{L}$'s and~$\mathcal{M}$. In turn these quantities may be calculated by specifying the following data:

\begin{itemize}
\item 
The two-point function $\mathcal{A}$
\eqref{eq: derivative-twopoint-distributions} in the field initial state $\ket{0}$ under consideration.
\item The trajectory followed by the detectors in question.
\item The switching functions that determine the duration of the interaction.
\end{itemize}

Sections \ref{sec:wightman_function} and \ref{sec:Trajectories_and_Switching_Function} describe the two-point function~$\mathcal{A}$, the trajectories, and the switching functions used in our problem.

Finally with $\rho\ts{AB}$ at hand, one may use one's favorite measure of entanglement to calculate the entanglement harvested. We review the measures used in our case in section~\ref{sec:Measures_of_Entanglement}.

\subsection{The Two-point Function}\label{sec:wightman_function}

We are interested in 
entanglement harvesting in the post-quench part of the quench spacetime, assuming that the field $\phi$ was prepared in the Minkowski vacuum in the pre-quench part. To compute this, we need the post-quench Wightman function, given by \cite{Louko_2018}
\begin{align}
W(\sx, \sx') &=
\bra{0 }{\hat \phi(\sx)  \hat \phi(\sx')}\ket{0} 
\nonumber\\ 
&= W_0(\eta - y, \eta' - y') + W_0(\eta + y, \eta' + y')
\label{eq:wightman-sum}
\end{align}
where 
\begin{align}
W_0\left(z, z'\right)
& = - \frac{\ii}{8} \sgn(z-z')
-\frac{1}{4 \pi} \ln \! \left[\sinh \! \left(\frac{|z - z'|}{2}\right)\right]
\notag
\\[1ex]
& \hspace{3ex}
-\frac{1}{4 \pi} \ln \! \left[\cosh \! \left(\frac{z+z^{\prime}}{2}\right)\right]. 
\label{eq:wightman-noeps}
\end{align}
While $W(\sx, \sx')$ is defined only up to an additive real-valued constant, the choice  made in \eqref{eq:wightman-noeps} for this constant does not lose generality, as it drops out of the two-point function $\mathcal{A}$~\eqref{eq: derivative-twopoint-distributions}. 
An alternative way to write \eqref{eq:wightman-noeps} is
\begin{align}
W_0\left(z, z'\right)
& = - \frac{\ii}{8}
-\frac{1}{4 \pi} \ln \! \left[\sinh \! \left(\frac{z - z' - \ii \varepsilon}{2}\right)\right]
\notag
\\[1ex]
& \hspace{3ex}
-\frac{1}{4 \pi} \ln \! \left[\cosh \! \left(\frac{z+z^{\prime}}{2}\right)\right], 
\label{eq:wightman}
\end{align}
where 
the limit $\varepsilon \to 0_+$ is understood, the logarithm in the second term is real for $z-z'>0$, and the branch of the logarithm for $z-z'<0$ is determined by analytic continuation. 

We shall use \eqref{eq:wightman} to compute 
the two-point function $\mathcal{A}$~\eqref{eq: derivative-twopoint-distributions}. 
The advantage of \eqref{eq:wightman} over \eqref{eq:wightman-noeps} is that the derivatives of \eqref{eq:wightman-noeps} would give for $\mathcal{A}$ an expression with distributional singularities, whereas 
\eqref{eq:wightman} allows us to bypass distributional techniques by keeping $\varepsilon >0$ during intermediate numerical steps. This, provided the computations remain accurate for sufficiently small~$\varepsilon$.

\subsection{Trajectories and Switching Functions}\label{sec:Trajectories_and_Switching_Function}

We consider detector trajectories generated by the timelike Killing vector~$\partial_\eta$. 
In the coordinates $(\eta,\chi)$, these trajectories read
\begin{equation}
X^\mu(\tau) = (\eta(\tau), \,\chi(\tau)) = \left(\frac{\tau}{\sqrt{\chi_0^2 + b^2}}\,,\, \,\chi_0\right) , 
\label{eq:traj-chicoord}
\end{equation}
where $\chi_0$ is a constant. Even though it also parametrizes its acceleration, we refer to $\chi_0$ as the position of the detector. 
In the coordinates $(\eta,y)$, the trajectory reads 
\begin{equation}
	X^\mu(\tau) = (\eta(\tau), y(\tau)) = \left(\frac{\tau}{b\cosh(y_0)}\,,\, y_0\right) , 
\label{eq:traj-ycoord}
\end{equation}
where $\chi_0 = b \sinh y_0$. 
With the additive constant in $\tau$ chosen as in \eqref{eq:traj-chicoord} and~\eqref{eq:traj-ycoord}, 
the quench occurs at $\tau=0$, and the detector in the post-quench spacetime has $\tau>0$. 

For the switching, we adopt
\begin{equation}\label{eq:switching_function}
\zeta(\tau) = \begin{cases}
\cos^4 \! \left(\pi\frac{\tau - \tau_c}{\sigma}\right) & \text{for } |\tau - \tau_c| < \frac{\sigma}{2}\\
0 & \text{otherwise}
\end{cases}
\end{equation}
where the positive parameter $\sigma$ is the duration of the interaction and $\tau_c$ is the value of $\tau$ at the mid-point of the interaction. We assume $\tau_c>\sigma/2$, so that the detector operates only in the post-quench spacetime. 
$\zeta(\tau)$ is $C^3$ at $\tau=\pm \sigma/2$ and $C^\infty$ elsewhere. 
Trajectory pairs that we consider are shown in the spacetime diagram of Figure~\ref{fig:swept_trajectories}, showing the support of the switching functions.

\subsection{Measures of Correlation}\label{sec:Measures_of_Entanglement}

The two measures of correlation of greatest interest are  concurrence \cite{Wotters1998entanglementmeasure} and mutual information \cite{nielsen2000quantum}.  For a two qubit density matrix of the form \eqref{eq:reduced_density_matrix} 
the concurrence is \cite{Horodecki996separable,smith2016topology}
\begin{align}\label{concur}
    \mathcal{C} [\rho\ts{AB}] 
    = 2\max\{0,\, |\mathcal{M}| - \sqrt{\mathcal{L}\ts{AA} \mathcal{L}\ts{BB} }\}+O(\lambda^4)
\end{align}
to lowest order in the coupling. For entanglement harvesting the 
relation \eqref{concur} is quite intuitive: entanglement can be extracted when the nonlocal quantity $|\mathcal{M}|$ is greater than the local `noise' contribution $\sqrt{ \mathcal{L}\ts{AA} \mathcal{L}\ts{BB} }$ of each individual detector.  
 
Another quantity of interest is the mutual information \cite{nielsen2000quantum}
\begin{align}
    I[\rho\ts{AB} ]
    \coloneqq S[\rho\ts{A} ]+S[\rho\ts{B} ]- S[\rho\ts{AB} ],
\end{align}
which is a measure of how much general correlation (including classical) is extracted by the detector.  The quantity $ S[ \rho ]\coloneqq-\Tr[ \rho \ln \rho ]$ is the von Neumann entropy. For the density matrix \eqref{eq:reduced_density_matrix} we have \cite{pozas2015harvesting}
\begin{align}\label{mutinf}
    I[\rho\ts{AB} ] 
    &= \mathcal{L}_+\ln \mathcal{L}_+ 
    + \mathcal{L}_- \ln \mathcal{L}_- \notag \\
    &- \mathcal{L}\ts{AA} \ln \mathcal{L}\ts{AA}
    - \mathcal{L}\ts{BB} \ln \mathcal{L}\ts{BB} 
    + O(\lambda^4), 
\end{align}
where 
\begin{align}
     \mathcal{L}_{\pm} 
     &= \frac{1}{2} 
     \kako{
        \mathcal{L}\ts{AA}
        + \mathcal{L}\ts{BB} 
        \pm 
        \sqrt{
            ( \mathcal{L}\ts{AA} - \mathcal{L}\ts{BB} )^2 
            + 4| \mathcal{L}\ts{AB} |^2
        }
     }.
\end{align}
If the two detectors have nonzero mutual information but vanishing concurrence then their correlations must be either classical correlation or non-entangling (such as quantum discord) \cite{Zurek2001discord,Henderson2001correlations}.  Note that $ I[\rho\ts{AB} ]=0$ 
if $\mathcal{L}\ts{AB}=0$.

\section{Setup of the Calculation}\label{sec:setup_of_calculation}

The following details the values of the parameters for which we calculated the entanglement harvested. We performed both concurrence and mutual information calculations. Henceforth, all dimensionful quantities are expressed in units of the width $\sigma$ of the switching function \eqref{eq:switching_function}.

\subsection{Setup for Concurrence Calculation}

We calculate the concurrence accumulated between the detectors, at different values of their energy gap, and at different separations between the trajectories. More specifically, we sweep over a range of values of the energy gap $\Omega$ and of the position of detector $B$, $\chi_{B}$, at a fixed value  $\chi_{A}$ of the initial position of detector $A$. The initial position of detector A was set to $\chi_{A} =  2\sigma$, while the ranges of $ \Omega $ and $ \chi_{B} $ are given in Table \ref{tab:sweeping_ranges}. The positions of the detectors were chosen so that all throughout the interaction the detectors remain spacelike.

\begin{table}[H]
\centering
\begin{tabular}{c|c|c}
            &  Min Value    &  Max Value  \\ \hline
$\Omega$    & $10/\sigma$   & $20/\sigma$ \\ \hline
$\chi_{B0}$ & $3.05\sigma$  & $3.5\sigma$ \\ \hline
$y_{B0}$    & $2.51$        & $2.64$      \\ \hline
\end{tabular}
\caption{Range of values over which we calculated the concurrence, expressed in terms of the duration of the interaction, $\sigma$.}
\label{tab:sweeping_ranges}
\end{table}

Furthermore, to study the influence of the pulse of the SET on the entanglement harvested, we fixed the centers of the switching functions, $\eta_c$, in three different scenarios: before, during and after the pulse. The specific values of the centers of the switching functions are shown in Table~\ref{tab:switching_funcs_centers}.

\begin{table}[H]
\centering
\begin{tabular}{c|c|c|c}
	& Before the Pulse & During the Pulse & After the Pulse \\
	\hline
	$\eta_c$ & $1.05$ & $2.09$ & $8.38$ \\
\end{tabular}
\caption{Values   of the $\eta$ coordinate of the center of the switching function, expressed in terms of the duration of the interaction, $\sigma$.}
\label{tab:switching_funcs_centers}
\end{table}

\begin{figure}[!h]
\centering
\includegraphics[width=1.0\linewidth]{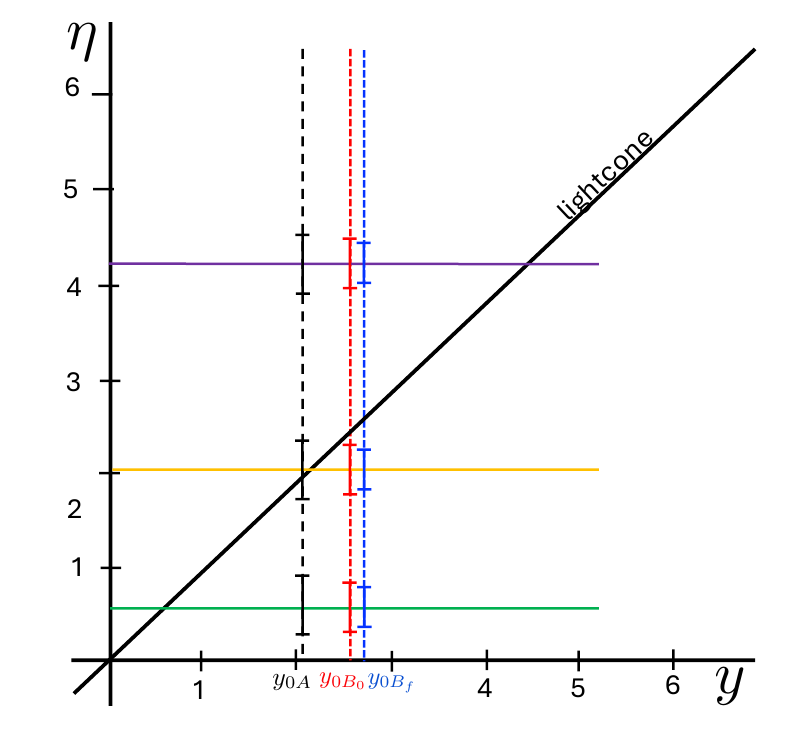}
\caption[]{Schematic image of the trajectories swept over to calculate their concurrence. The trajectory $X_A$ was kept at fixed $y_{0A}$, while that for $X_B$ was swept from $y_{0B_0}$ (red) to $y_{0B_f}$
(blue). The supports of the switching functions were studied in three separate cases, \textit{above} (purple), \textit{at} (orange), and \textit{below} (green) the lightcone. These supports are shown with bolder lines.}
	\label{fig:swept_trajectories}
\end{figure}

A schematic representation of the trajectories swept, and the different conditions studied for the switching function are shown in Figure~\ref{fig:swept_trajectories}.

Some other parameter values of note  are:
\begin{itemize}
\item The parameter that controls the deformation of the quench spacetime in~\eqref{eq:quench_metric}: $b=\tfrac12\sigma$. 
\item For the numerical calculations, we set the width of the support of the switching function $\sigma$ to~$1$.

\end{itemize}

The coupling constant $\lambda$ enters $\mathcal{L}_{ij}$ \eqref{eq:Lij} and $\mathcal{M}$ \eqref{eq:nonlocal M}
only as the overall multiplicative factor~$\lambda^2$. The concurrence \eqref{concur} is hence proportional to~$\lambda^2$. The concurrence plots in Figures 
\ref{fig:concurrence_below_pulse}--\ref{fig:rindler_reference} give the concurrence divided by $\lambda^2$ and are hence independent of~$\lambda$. The mutual information \eqref{mutinf} is by construction independent of~$\lambda$. The value of $\lambda$ does therefore not appear in our results.

\subsection{Setup for Mutual Information Calculation}

We compute the mutual information \eqref{mutinf} as a function of the energy gap $\Omega$ and the position of detector B, $\chi_B$, in the same three scenarios as before: before, at, and after the pulse of the SET; a depiction of the setting is again given in Figure \ref{fig:swept_trajectories}. 

The ranges over which the energy gap $\Omega$ and the position of detector $B$ were varied were adjusted from those used in the concurrence calculations, in order to focus on the region of the plot where the mutual information exhibited its most salient features. The specific values used are listed in Table \ref{tab:mutual_info_sweeping_ranges}.

\begin{table}[H]
\centering
\begin{tabular}{c|c|c}
            &  Min Value    &  Max Value   \\ \hline
$\Omega$    & $-5/\sigma$   &  $15/\sigma$ \\ \hline
$\chi_{B0}$ & $3.05\sigma$  & $4.0\sigma$  \\ \hline
$y_{B0}$    & $2.51$        & $2.78$       \\ \hline
\end{tabular}
\caption{Range of values over which we calculated the mutual information, expressed in terms of the duration of the interaction, $\sigma$.}
\label{tab:mutual_info_sweeping_ranges}
\end{table}

\section{Results}\label{sec:results}

\subsection{Concurrence}

Density plots of the concurrence calculated as a function of the energy gap, $\Omega$, and the initial position of detector~$B$, $\chi_{B}$, are shown in figures \ref{fig:concurrence_below_pulse}, \ref{fig:concurrence_at_pulse} and \ref{fig:concurrence_above_pulse}. 
Figure \ref{fig:concurrence_below_pulse} corresponds to the case where the detectors interact before the SET pulse, Figure \ref{fig:concurrence_at_pulse} to the case where they interact at the pulse, and Figure \ref{fig:concurrence_above_pulse} corresponds to the case where they interact after the pulse. 

For comparison, Figure \ref{fig:rindler_reference} shows
the concurrence for a pair of detectors in Rindler spacetime in the Minkowski vacuum, with the corresponding parameter values.

From the figures we observe that the entanglement harvested when the two detectors operate before the SET pulse or after the SET pulse is very similar to the entanglement harvested in Rindler spacetime. 
When the detectors operate at the SET pulse, however, the entanglement harvested shows a small but distinct increase. 

In all the figures we see that the entanglement harvested peaks with an energy gap between $13/\sigma$ and $16/\sigma$ depending on the separation between the detectors. Also, for all values of $ \Omega $ the concurrence decreases as we separate the detectors.

\subsection{Mutual Information}

The density plots of the mutual information as a function of the energy gap, $\Omega$, and the initial position of detector~$B$, $\chi_{B0}$ are shown in figures \ref{fig:mutual_information_below_pulse}, \ref{fig:mutual_information_at_pulse}, and~\ref{fig:mutual_information_above_pulse}, 
corresponding to the cases of below, at, and above the SET pulse, respectively.

When the two detectors are operating before the SET pulse or after the SET pulse, the plots are closely similar. When the detectors operate at the SET pulse, however, the mutual information shows a small but distinct increase.

All three figures show a similar dependence of the mutual information in terms of the energy gap, $\Omega$, and the position of detector $B$, $\chi_{B0}$. 
Mutual information peaks around an energy gap of $4/\sigma$ for all values of~$\chi_{B0}$. Also, for all values of $ \Omega $ plotted, the mutual information decreases as we separate the detectors.

\section{Conclusions}\label{sec:conclusions}

Motivated by the proposal of \cite{Rodriguez_Laguna_2017,Kosior_2018} 
to simulate the Unruh effect experimentally with ultracold fermionic atoms in an optical lattice, by a rapid quench in the dispersion relation, we have analysed the entanglement of a scalar field in a similar $(1 + 1)$-dimensional continuum quench spacetime, and the harvesting of this entanglement by a pair of Unruh-DeWitt detectors. We found that the concurrence and mutual information harvested by the pair of detectors are indeed qualitatively similar to those in Rindler spacetime, but they are slightly yet distinctly higher when the energy pulse created by the quench crosses the detectors' worldlines.

Our results strongly suggest that entanglement is present also in the optical lattice setup of~\cite{Rodriguez_Laguna_2017,Kosior_2018}, and they raise the possibility that entanglement harvesting could be experimentally realised there. 
There are certainly quantitative differences between the optical lattice setup of \cite{Rodriguez_Laguna_2017,Kosior_2018} and the system that we have analysed: these differences include spatial discreteness versus spatial continuum, spacetime dimension 2+1 versus spacetime dimension 1+1, and a fermionic effective field versus a boson field; they also include our employment of a derivative-coupled UDW detector versus any entanglement-extraction mechanism that may be employable for a quantum dot in a fermionic optical lattice. All that being said, the geometric similarities are compelling. 
We see our results as a further rationale to realise the experimental proposal of~\cite{Rodriguez_Laguna_2017,Kosior_2018}.


\begin{widetext}

\begin{figure}[H]
	\centering
	\includegraphics[width=0.85\linewidth]{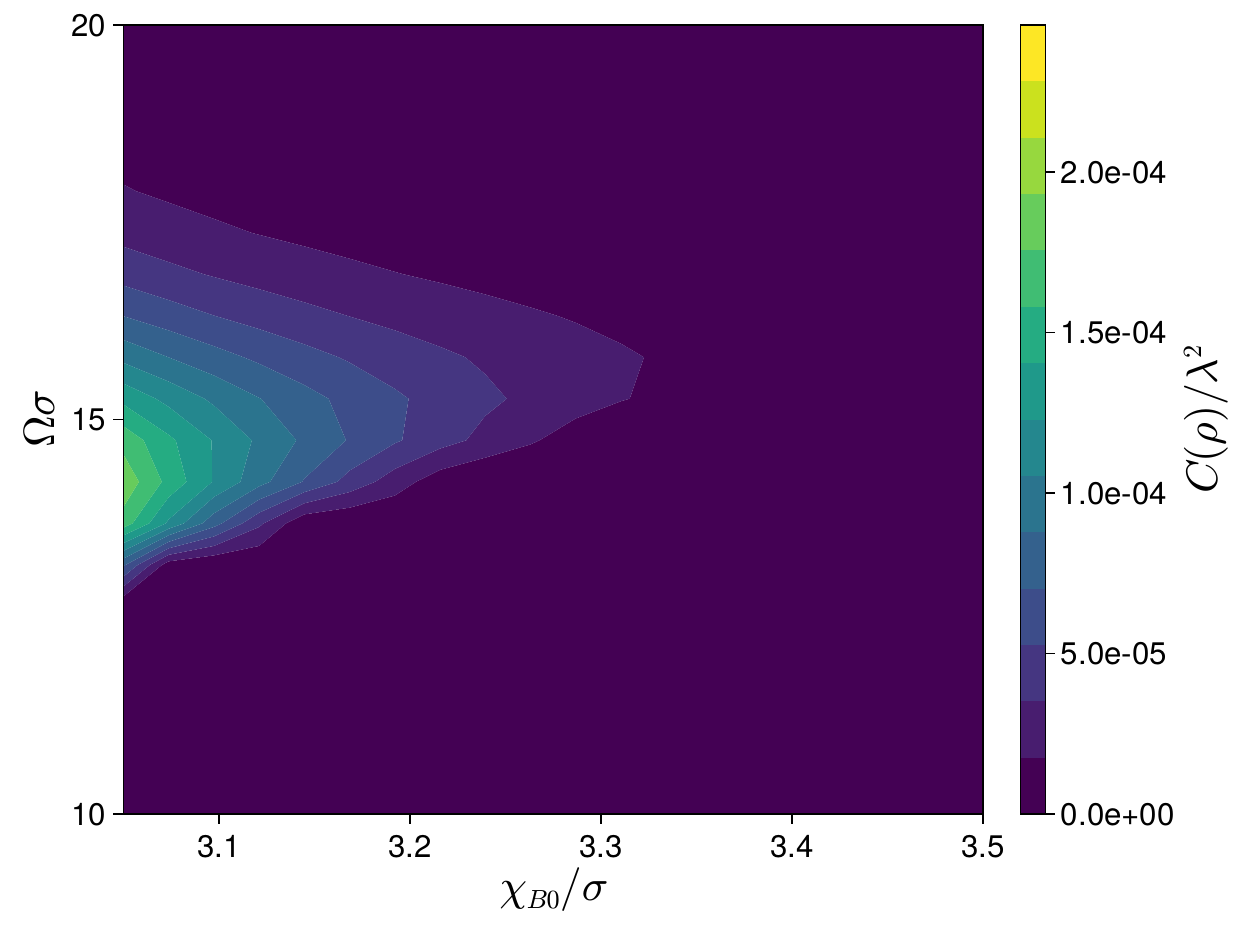}
	\caption[]{Concurrence obtained by detectors interacting \emph{below} the SET pulse, as a function of $\Omega$ and initial position of detector B, $\chi_{B0}$.}
	\label{fig:concurrence_below_pulse}

	\centering
	\includegraphics[width=0.85\linewidth]{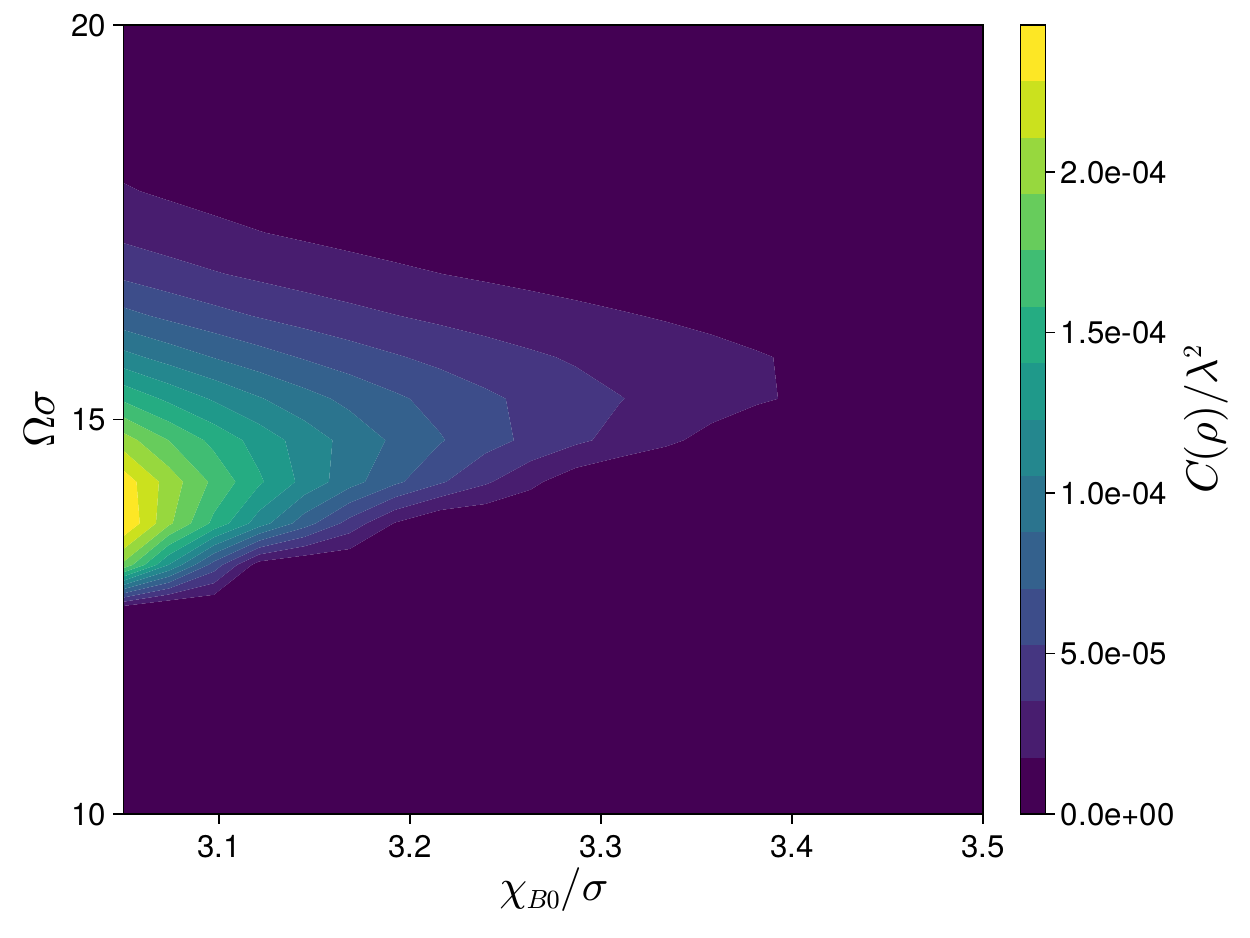}
	\caption[]{Concurrence obtained by detectors interacting \emph{at} the SET pulse, as a function of $\Omega$ and initial position of detector B, $\chi_{B0}$.}
	\label{fig:concurrence_at_pulse}
\end{figure}
\begin{figure}[H]
	\centering
	\includegraphics[width=0.85\linewidth]{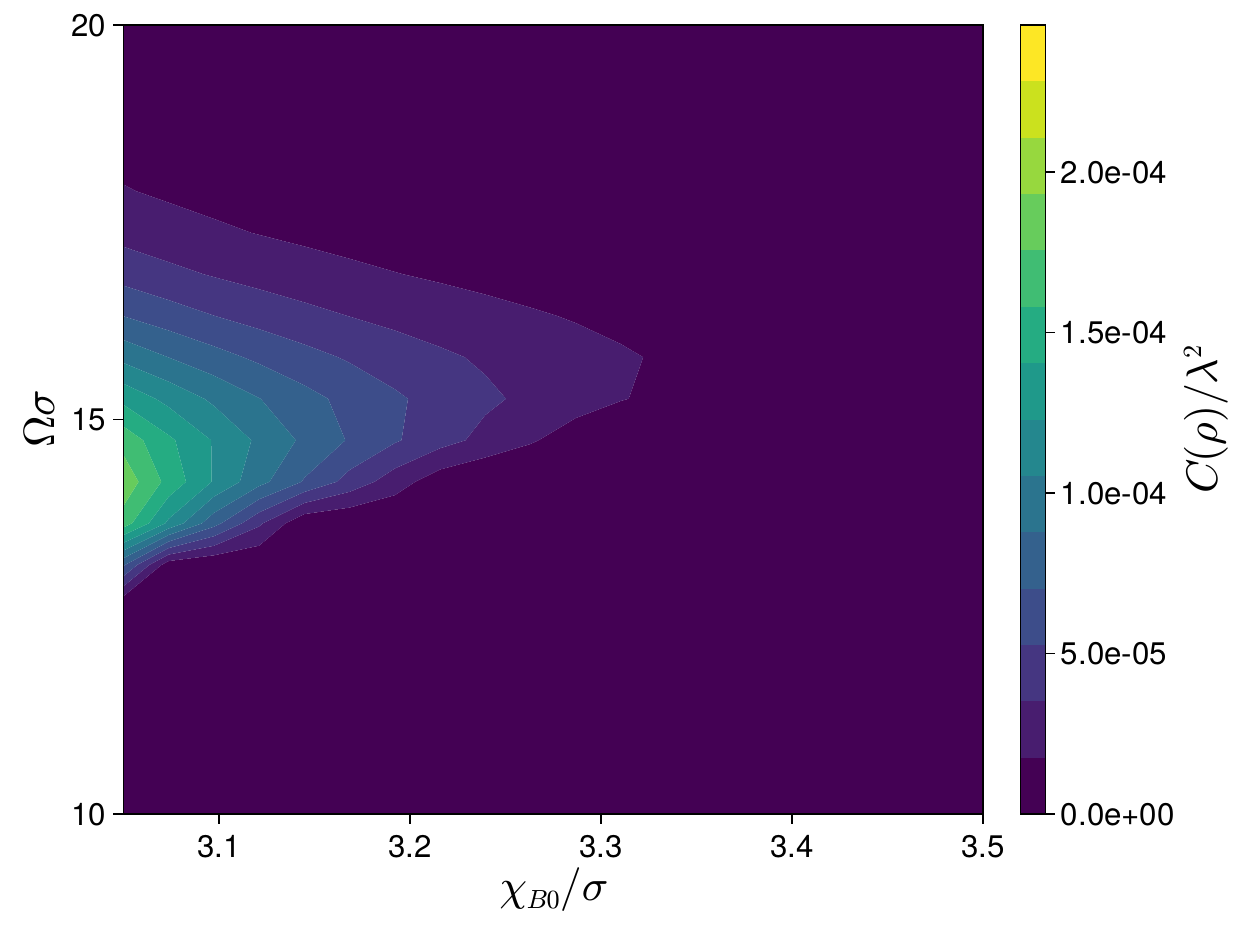}
	\caption[]{Concurrence obtained by detectors interacting \emph{above} the SET pulse, as a function of $\Omega$ and initial position of detector B, $\chi_{B0}$.}
	\label{fig:concurrence_above_pulse}
	
	\centering
	\includegraphics[width=0.85\linewidth]{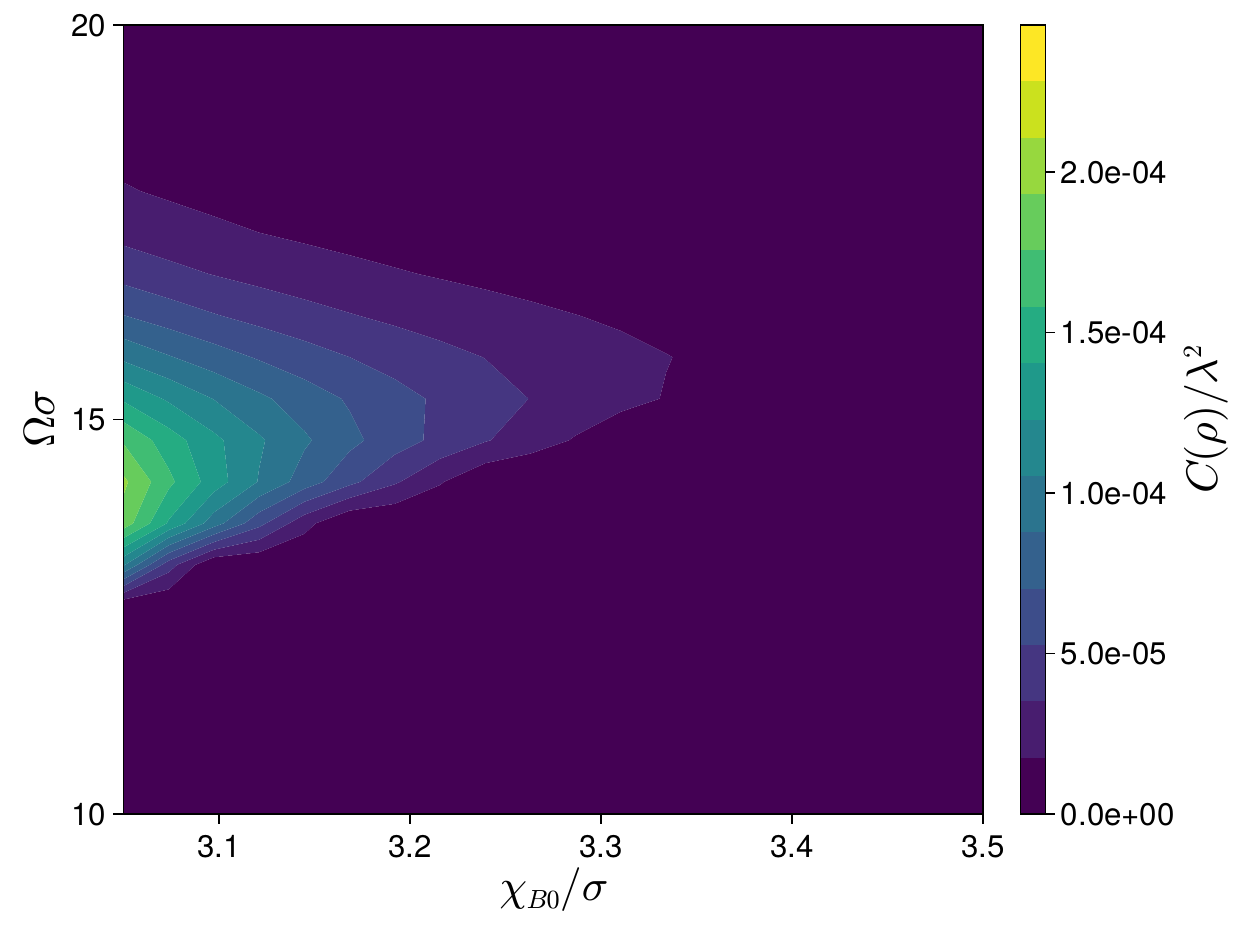}
	\caption[]{Concurrence obtained by accelerated detectors in flat 2d Minkowski, as a function of $\Omega \sigma$ and initial position/acceleration of detector B, $\chi_{B0}\sigma$.}
	\label{fig:rindler_reference}
\end{figure}

\begin{figure}[H]
	\centering
	\includegraphics[width=0.85\linewidth]{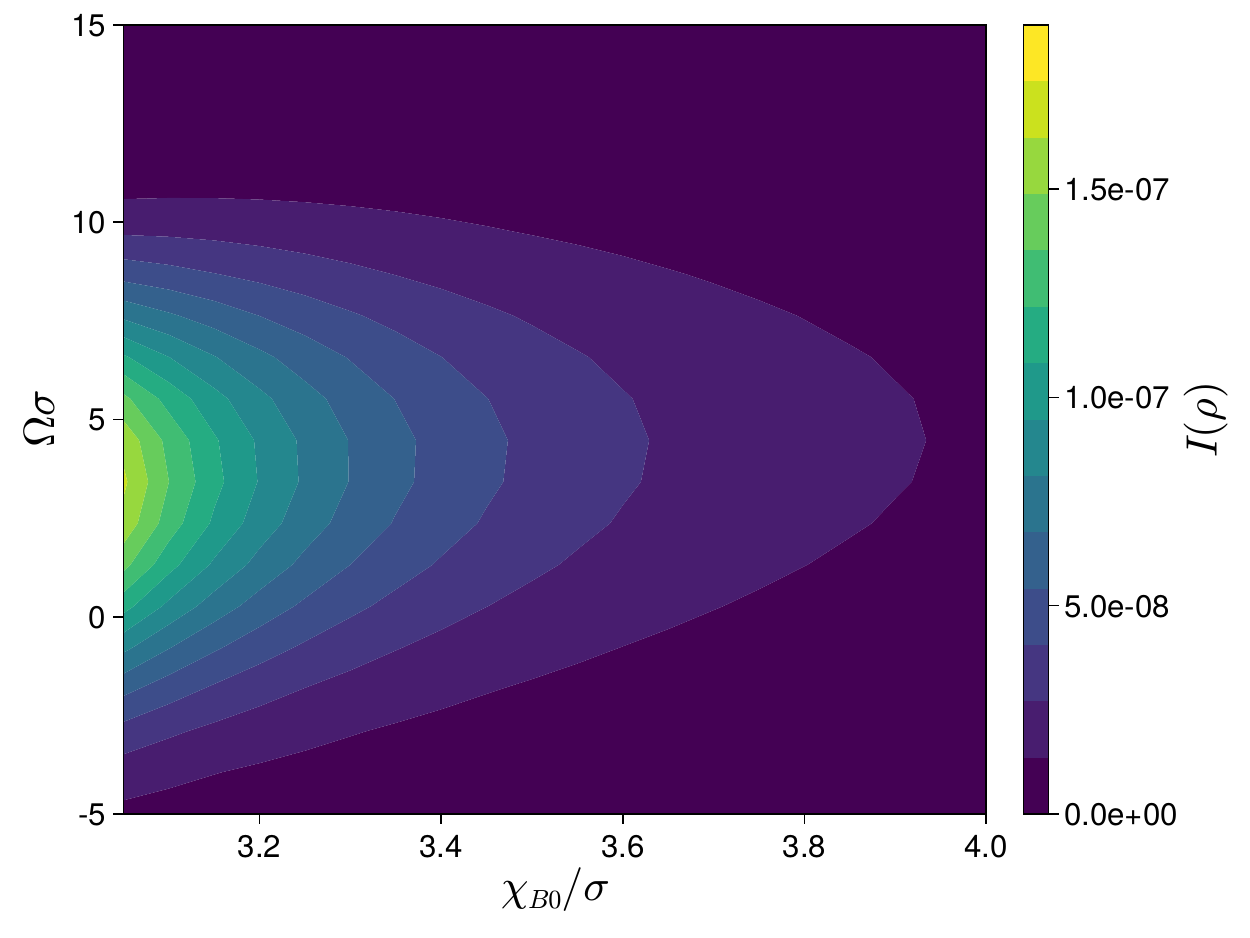}
	\caption[]{Mutual Information obtained by detectors interacting \emph{below} the SET pulse, as a function of $\Omega$ and initial position of detector B, $\chi_{B0}$.}
	\label{fig:mutual_information_below_pulse}
	
	\centering
	\includegraphics[width=0.85\linewidth]{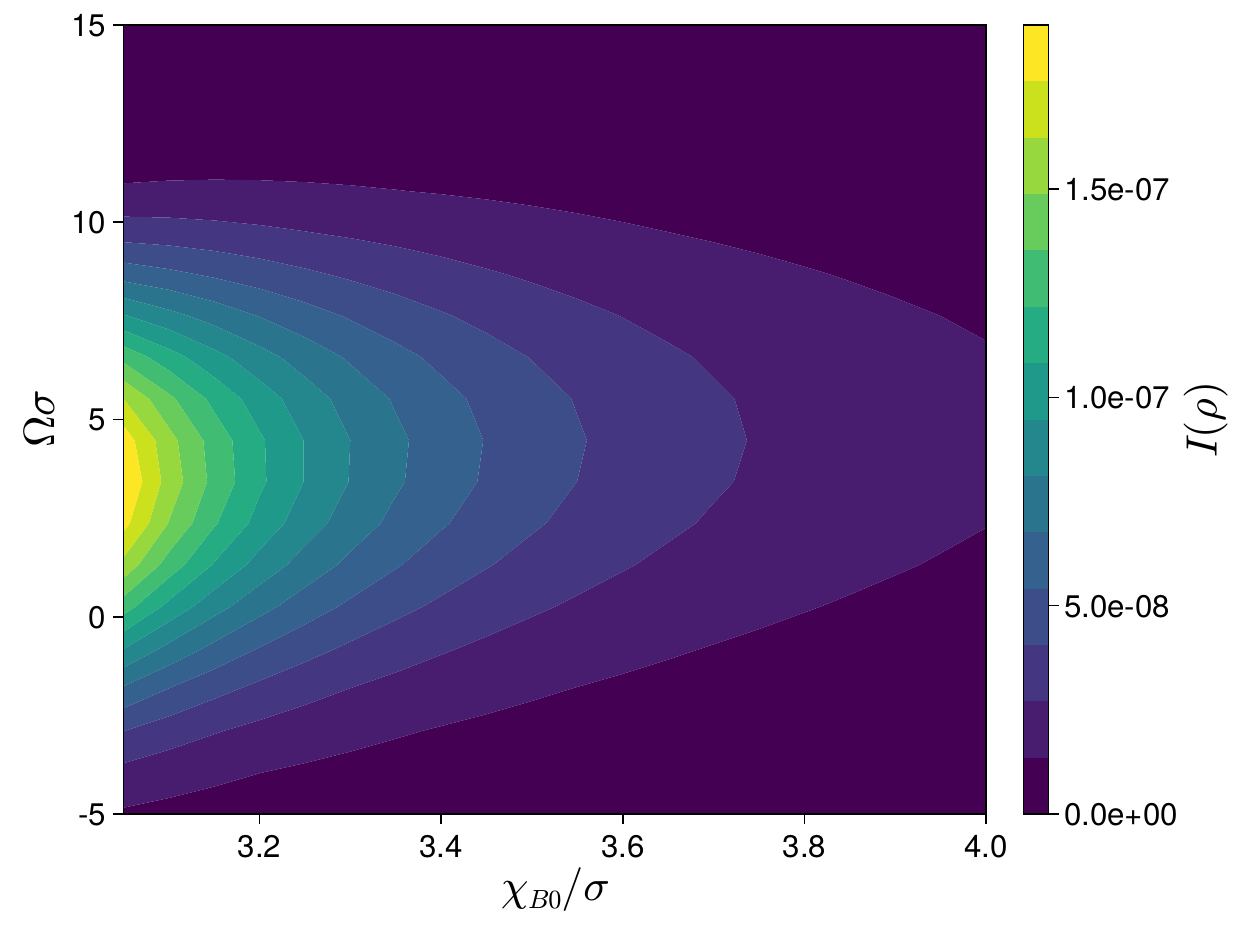}
	\caption[]{Mutual Information obtained by detectors interacting \emph{at} the SET pulse, as a function of $\Omega$ and initial position of detector B, $\chi_{B0}$.}
	\label{fig:mutual_information_at_pulse}
\end{figure}

\begin{figure}[H]
    \centering
	\includegraphics[width=0.85\linewidth]{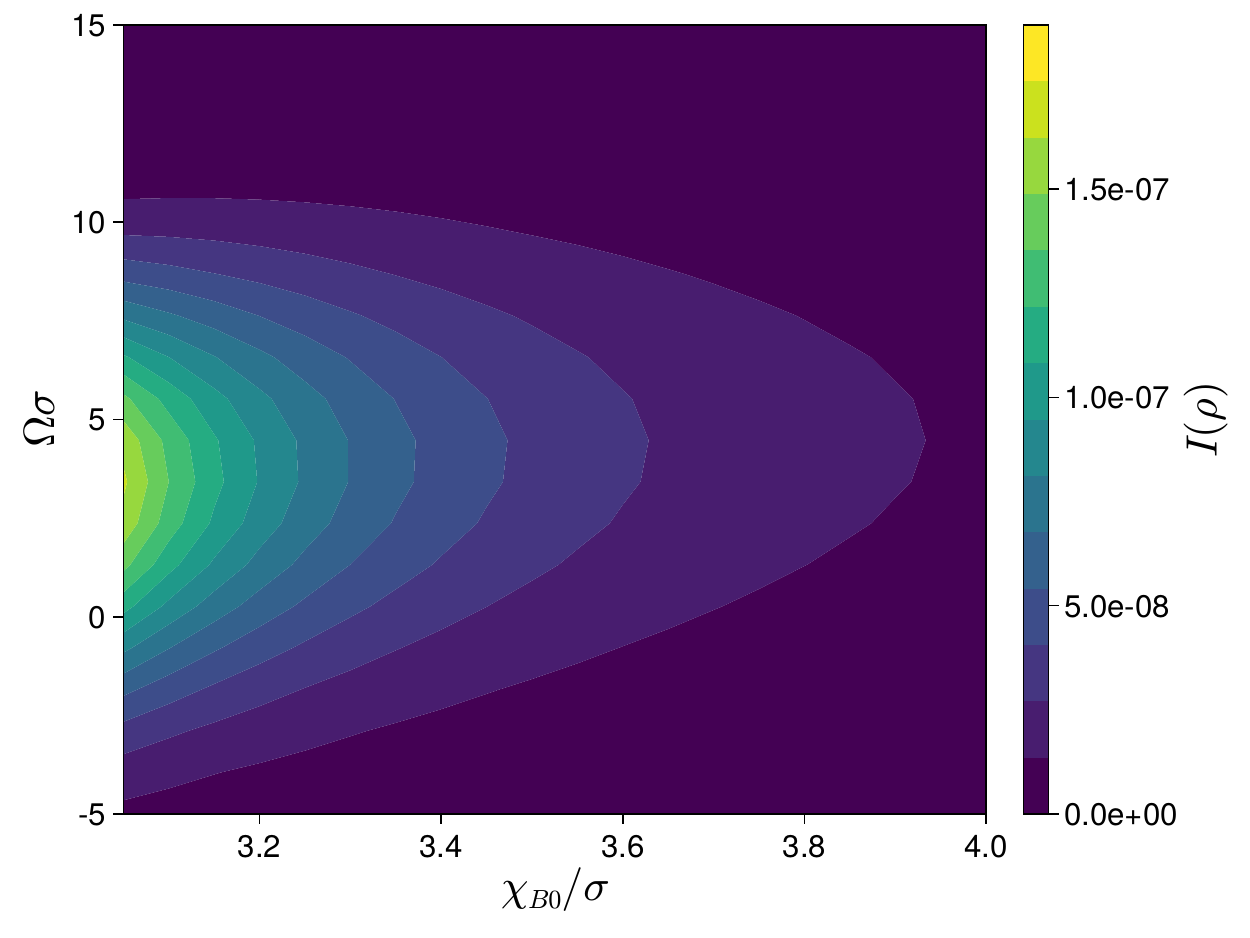}
	\caption[]{Mutual Information obtained by detectors interacting \emph{above} the SET pulse, as a function of $\Omega$ and initial position of detector B, $\chi_{B0}$.}
	\label{fig:mutual_information_above_pulse}
\end{figure}

\end{widetext}

\section*{Acknowledgements}
We thank an anonymous referee for helpful presentational comments. 
This work was supported in part by the Natural Sciences and Engineering Research Council of Canada. The work of JL was supported by United Kingdom Research and Innovation Science and Technology Facilities Council [grant numbers ST/S002227/1, ST/T006900/1 and ST/Y004523/1]. 
For the purpose of open access, the authors have applied a CC BY public copyright licence to any Author Accepted Manuscript version arising. 

\appendix

\section{About the Code used for the Calculations}\label{sec:about_code}

The code used for the calculation is publicly available~\cite{code-available-github}. 
It is designed to be as reusable as possible for future entanglement harvesting projects. For any questions about the structure or operation of the code, please contact the first author at a22lopez@uwaterloo.ca or alopezraven@perimeterinstitute.ca.

The code basically consists of a series of functions that lead up to the calculation of expressions \eqref{eq:Lij}
and \eqref{eq:nonlocal M} using numerical integration, after which the density matrix may be determined through~\eqref{eq:reduced_density_matrix}, from which the concurrence can be immediately calculated.

\subsection{Validation of the Code}

Given that the entanglement harvesting results calculated in this project have no analytical results or experimental data against which to compare/verify them, to guarantee a certain degree of reliability of our results, we used the same code to calculate properties of other spacetimes which have been previously calculated in the literature so as to make sure we could reproduce them, therefore providing some trust in our numerical calculations. More specifically we:

\begin{enumerate}
\item Calculated the reduced density matrix elements of an inertial trajectory in 4d flat spacetime, with Gaussian switching function, coupled to a massless free scalar; we plotted these results against the separation of the detectors and against their energy gap, and compared these results to the ones calculated analytically in~\cite{Alex_Thesis}. Our numerical calculations perfectly matched the analytical results found in~\cite{Alex_Thesis}. 

\item We calculated the entanglement harvested by accelerated detectors in 4d flat spacetime and compared our results to those obtained also numerically in~\cite{Accelerated_detectors}. Again, we found perfect agreement with their results.

\item We calculated the entanglement harvested in the quench spacetime in the limit of $ b $ going to zero (i.e.\ the limit in which one should recover accelerated trajectories in flat spacetime) with fixed~$\sigma$, and we calculated the entanglement harvest by accelerated trajectories in flat spacetime finding that the two coincided.
\end{enumerate}

\bibliography{refs.bib}
\bibliographystyle{JHEP.bst}

\end{document}